\documentclass{jfm}

\begin{document}

\newtheorem{lemma}{Lemma}
\newtheorem{corollary}{Corollary}

\shorttitle{Dimensional Reduction of DSS} 
\shortauthor{A. Allawala et al} 

\title{Dimensional Reduction of Direct Statistical Simulation}

\author
 {
 Altan Allawala\aff{1,2},
  S. M. Tobias\aff{3}
  \and 
  J. B. Marston\aff{1}
   \corresp{\email{marston@brown.edu}}
  }

\affiliation
{
\aff{1}
Department of Physics, Brown University, Providence, Rhode Island 02912-1893, USA
\aff{2}
JPMorgan Chase \& Co., 237 Park Ave., New York, New York 10016
\aff{3}
Department of Applied Mathematics, University of Leeds, Leeds LS2 9JT, UK
}

\maketitle

\begin{abstract}
Direct Statistical Simulation (DSS) solves the equations of motion for the statistics of turbulent flows in place of the traditional route of accumulating statistics by Direct Numerical Simulation (DNS).  That low-order statistics usually evolve slowly compared with  instantaneous dynamics is one important advantage of DSS.  Depending on the symmetry of the problem and the choice of averaging operation, however, DSS is usually more expensive computationally than DNS because even low-order statistics typically have higher dimension than the underlying fields.     Here we show that it is possible to go much further by using a form of unsupervised learning, Proper Orthogonal Decomposition (POD), to address the ``curse of dimensionality.''   We apply POD directly to DSS in the form of expansions in the equal-time cumulants to second order (CE2).  We explore two averaging operations (zonal and ensemble) and test the approach on two idealized barotropic models of fluid on a rotating sphere (a jet that relaxes deterministically towards an unstable profile, and a stochastically-driven flow that spontaneously organizes into jets).  Order-of-magnitude savings in computational cost are obtained in the reduced basis, potentially enabling access to parameter regimes beyond the reach of DNS.
\end{abstract}

\section{Introduction}
\label{sec:Introduction}

Statistical descriptions are appropriate for turbulent flows.  In nature such flows are rarely homogeneous and isotropic; instead they typically exhibit rich correlations that reflect the presence of coherent structures, with statistics that evolve only slowly in time, or not at all. 
The large range of spatial and temporal scales spanned by turbulent flows often makes their Direct Numerical Simulation (DNS) computationally prohibitive \citep{bauer2015quiet,tobias2019}.  Fifty years ago Lorenz pointed to an alternative approach that
``consists of deriving a new system of equations whose
unknowns are the statistics themselves'' \citep{lorenz1967nature}.  Such Direct Statistical Simulation (DSS) has seen, in recent years, a number of successful applications \citep{Farrell:2007fq,marston65conover,marston2010statistics,tobias2011astrophysical,Marston:2012co,tobias2013direct,Constantinou:2013fh,marston2014direct,Laurie:2014dn}.

Many well-established approaches can be considered instances of DSS.
For example the probability distribution function can be obtained from the Fokker-Planck equation 
by numerical methods, but this is limited to dynamical systems of at most a few dimensions  \citep{bergman1992robust,pichler2013numerical,von2000calculation,naess1994response,kumar2006solution,Allawala:2016hx}.
Another form of DSS, Large Deviation Theory \citep{largedeviationtheory,Bouchet:2009cl,Laurie:2015co,Bouchet:2018er},
focuses on extreme or rare events.  Other methods such as those developed by Kolmogorov \citep{batchelor1947kolmogoroff},
Kraichnan \citep{frisch1995turbulence},
and others \citep{legras1980turbulent,holloway1977stochastic,huang2001anisotropic}
provide an approximate description of some statistical properties of turbulent flows
but assume homogeneity and usually isotropy. Many flows in geophysics and astrophysics 
spontaneously develop features such as coherent vortices and zonal banding \citep{marston2014direct,Skitka:2020}. Furthermore, in engineering applications turbulence often interacts with non-trivial mean flows \citep[see e.g.][]{barkley2016}. Therefore,
any statistical method that can appropriately treat such systems needs
to respect such asymmetries. One such scheme of DSS that meets these requirements is that of low-order expansions in equal-time (but spatially nonlocal) cumulants.  Since low-order statistics
are spatially smoother than the corresponding dynamical fields (or instantaneous flow), the approach can capture the macroscopic features of turbulent flows using fewer degrees of freedom. An added benefit of such a cumulant expansion scheme is that the detailed time evolution of the flow is replaced by a description of the statistics of most interest.  The modes associated with the low-order statistics may be described by a fixed point or a slow manifold that can be quickly accessed. 

It is important to note that naive implementations of expansions in cumulants may be much more expensive computationally than full DNS because the second cumulant may have higher dimension than the underlying fields (depending on the symmetry of the problem and choice of averaging operation).  In this paper we investigate a reduced dimensionality method for DSS, based on a Proper Orthogonal Decomposition (POD) of the eigenvectors of the second moment.  This is a form of unsupervised learning, with training based upon full resolution simulations.  The equations of motion (EOMs) for the cumulants are rotated into a sub-basis formed by the eigenvectors of the second zonally-averaged moment after removal of eigenvectors with small eigenvalues.  We implement POD directly on the simplest non-trivial closure, one that goes to second order in an expansion of cumulants (CE2), of two model problems in fluid dynamics.

The rest of the paper is organized as follows.  In Section \ref{sec:CumulantExpansion} we introduce the two different types of cumulant expansions that are explored.  Although CE2 is often performed using a zonal average \citep{marston65conover,marston2014direct} (Section \ref{sub:ZonalAverage}), this has the drawback that scattering of eddies off non-zonal coherent structures such as vortices are neglected \citep{Tobias:2017im}.  We therefore also explore a variant of CE2 that is based upon an ensemble average \citep{bakas2011structural,bakas2013emergence,bakas2014theory,Allawala:2016hx} (Section \ref{sub:EnsembleAverage}). Although more accurate, ensemble-averaged cumulants have higher dimensionality compared with those based upon the zonal average; this is partly overcome with our POD method as discussed in Section \ref{sub:POD}.  In Section \ref{sec:Tests} both types of CE2 are evaluated against DNS which serves as the reference truth. We test the approaches on two different highly idealized barotropic models of planetary atmospheres on a spherical geodesic grid: A deterministic point jet relaxed toward an unstable profile (Section \ref{sub:DeterministicJet}), and a stochastically-forced jet (Section \ref{sub:StochasticJet}). The order of magnitude computational savings of DSS in a reduced basis with little loss to accuracy promises a fast and accurate alternative to accessing directly the low-order statistics of turbulent flows, and offers the possibility that flow regimes inaccessible to DNS will come within reach.  This speed-up is illustrated by a continuation in parameter space, keeping the POD basis fixed, in Section \ref{sec:Continuation}.  Section \ref{sec:Conclusion} concludes with some discussion.

\section{Cumulant Expansions}
\label{sec:CumulantExpansion}

We carry out a non-equilibrium statistical closure of the low order equal-time statistics of the flow \citep{marston2014direct,marston65conover}. The approach can be more easily understood by application to a simple toy model. We do so here by considering a barotropic (two-dimensional) fluid on a rotating sphere of unit radius, where relative vorticity evolves under the action of a bilinear Jacobian operator,
$J\left[A,~ B\right]=\hat{r}\cdot(\nabla A\times\nabla B)$,
a linear operator that contains frictional and hyperviscous terms 
$L\left[A\right] \equiv -\left[\kappa-\nu_3(\nabla^2+2)\nabla^4\right] A$,
and either a deterministic forcing term $F$, stochastic forcing $\eta$, or both.  
The EOM of the barotropic model is then given by:
\begin{equation}
\dot{\zeta} = J\left[\zeta+f,~ \psi\right] + L\left[\zeta\right] + F + \eta,
\label{eq:BarotropicEOM}
\end{equation}
where $\psi$ is the stream function, $\zeta=\nabla^2\psi$ is the
relative vorticity, $f=2\Omega\cos(\theta)$ is the Coriolis
parameter, $\theta$ is the co-latitude and $\phi$ is the azimuth
angle.  We set $\Omega = 2 \pi$ and thus the unit of time is the period of rotation, a day.  

The cumulant expansion may then be implemented by Reynolds decomposing
the relative vorticity into the sum of an average
vorticity field $\overline{\zeta}$ and a fluctuation about that average
$\zeta^{\prime}$ so that 
$\zeta(\theta,\phi)=\overline{\zeta}(\theta,\phi)+\zeta^{\prime}(\theta,\phi)$.
The choice of this averaging operation will be postponed until later, but it will be required to
satisfy the Reynolds averaging rules:
\begin{equation}
\overline{\zeta^{\prime}(\theta,\phi)}  =  0, \quad \quad
\overline{\overline{\zeta(\theta,\phi)}}  =  \overline{\zeta(\theta,\phi)}, \quad \quad
\overline{\overline{\zeta}\zeta} =  \overline{\zeta}\,\overline{\zeta}.
\end{equation}
The first three cumulants are centered moments, and on the surface of a sphere the first two read:
\begin{eqnarray}
c(\theta_1,\phi_1) &\equiv&  \overline{\zeta(\theta_1,\phi_1)},
\nonumber\\
c(\theta_1,\phi_1;\theta_2,\phi_2)  &\equiv&  \overline{\zeta^{\prime}(\theta_1,\phi_1)\zeta^{\prime}(\theta_2,\phi_2)}.
\label{cumulants}
\end{eqnarray}

Since the Jacobian couples the relative vorticity
and the stream function, it is useful to define their correlations
as auxiliary cumulants,
\begin{eqnarray}
p(\theta_1,\phi_1)  &\equiv&  \overline{\psi(\theta_1,\phi_1)},
\nonumber \\
p(\theta_1,\phi_1;\theta_2,\phi_2)  &\equiv&  \overline{\zeta^{\prime}(\theta_1,\phi_1)\psi^{\prime}(\theta_2,\phi_2)}.
\label{auxilliary}
\end{eqnarray}
The EOMs of the cumulants are derived by applying the
averaging operation to Equation~(\ref{eq:BarotropicEOM}). Refer to \citet{marston2014direct} for a detailed derivation. The first cumulant evolves as:
\begin{eqnarray}
\frac{\partial}{\partial t}c(\vec{\Omega}_1) &=&  
\int J_1\left[p(\vec{\Omega}_1,~ \vec{\Omega}_2),\delta(\vec{\Omega}_1 - \vec{\Omega}_2)\right] d\Omega_2
\nonumber \\
&+& J_1\left[c(\vec{\Omega}_1) + f(\theta_1),~ p(\vec{\Omega}_1)\right]
\nonumber \\
&+& L_1\left[c(\vec{\Omega}_1)\right] + F(\vec{\Omega}_1)
\label{eq:c1}
\end{eqnarray}
where the subscript $1$ on $J$ and $L$ indicates that these operators act upon the vector coordinate $\vec{\Omega}_1 \equiv (\theta_1, \phi_1)$
and $\delta(\vec{\Omega}_1 - \vec{\Omega}_2)$ is the two-dimensional Dirac functional.

As averages of the product of two fields do not generally equal the product of their separate averages,  the quadratic nonlinearity
leads to the well-known closure problem with the equation of motion
of the first cumulant depending on the second cumulant.
Likewise, the equation of motion for the second cumulant involves
the first, second and third cumulant. 
A closure should be performed at the lowest order possible; this may be achieved by decoupling the third
cumulant from the EOM of the second cumulant, known as the CE2 approximation:
\begin{equation}
\overline{\zeta^{\prime}(\vec{\Omega}_1)~ \zeta^{\prime}(\vec{\Omega}_2)~ \psi(\vec{\Omega_3})} 
\simeq  c(\vec{\Omega}_1,~\vec{\Omega}_2)~ p(\vec{\Omega}_3)\,\,\,\,\,{\rm (CE2)}.
\end{equation}
With this approximation the EOM for the second cumulant closes to give
\begin{eqnarray}
\frac{\partial}{\partial t} c(\vec{\Omega}_1, \vec{\Omega}_2) 
&=& 2 \big\{ J_1\left[ c(\vec{\Omega}_1) + f(\theta_1),~ p(\vec{\Omega}_2, \vec{\Omega}_1) \right] 
\nonumber \\
&+& 2  J_1\left[c(\vec{\Omega}_1, \vec{\Omega}_2),~ p(\vec{\Omega}_1)\right] 
\nonumber \\
&+& L_1\left[c(\vec{\Omega}_1,\vec{\Omega}_2)\right] \big\} + 2 \Gamma(\vec{\Omega}_1,\vec{\Omega}_2),
\label{eq:c2}
\end{eqnarray}
where the symmetrization operator $\left\{ \cdots\right\} $ performs
an average over all interchanges of the field points.
Here $\Gamma(\vec{\Omega}_1,\vec{\Omega}_2)$ is the covariance of 
the Gaussian stochastic forcing $\eta(\vec{\Omega},t)$ that is assumed to be $\delta$-correlated in time.
CE2 neglects
interaction between two fluctuations to produce another fluctuation because of the decoupling of the third and higher cumulants
\citep{herring1963investigation,schoeberl1984numerical}. Thus the
fluctuation-fluctuation scattering process is neglected, making it formally equivalent
to a quasi-linear approximation \citep{o2007recovery,herring1963investigation}.
This confers upon the CE2 approximation two attractive properties:
conservation up to quadratic order \citep{legras1980turbulent} (angular momentum, energy and enstrophy) in
the limit of no forcing or dissipation, and
physical realizability \citep{hanggi1980remark,kraichnan1980realizability,salmon1998lectures} with a positive-definite second cumulant.  
These properties ensure numerical stability.  Both the DNS and DSS performed here  are carried out in a basis of spherical harmonics (see \citet{marston2014direct})
with the spectral decomposition:
\begin{equation}
\zeta(\theta, \phi) = \sum_{\ell = 1}^L \sum_m^{|m| \leq \min(\ell, M)} \zeta_{\ell m}~ Y_\ell^m(\theta, \phi). 
\end{equation}
Spectral cutoffs $L$ and $M$ are specified below.\footnote{A program that implements the computations is available on the Apple Mac App Store at URL https://apps.apple.com/us/app/gcm/id592404494?mt=12.}

\subsection{Zonal Average}
\label{sub:ZonalAverage}

\begin{figure}
\centering{}\includegraphics[clip,width=1\columnwidth]{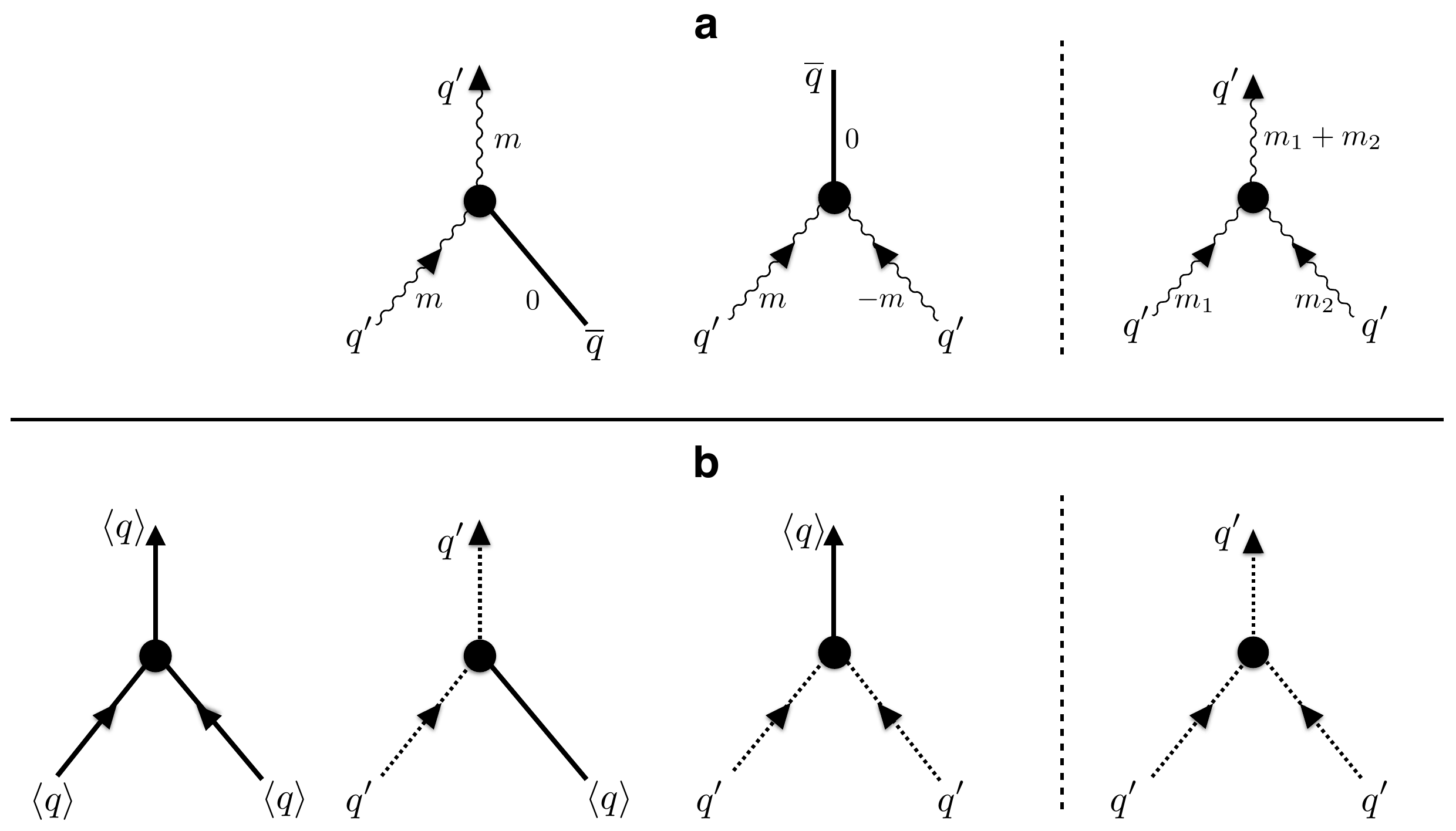}
\caption{(a) Retained (left) and discarded (right) triadic interactions in
zonal CE2.  Solid lines denote the amplitude of the zonal mean flow.  Wavy lines represent eddies
with zonal wavenumber $m$ that is unchanged by the interaction with the zonal mean flow.  
(b) Same as (a) but for ensemble CE2.  Solid lines indicate
the coherent part of the flow, and dashed lines represent the incoherent parts.  
The zonal wavenumber now generally changes when the coherent and/or incoherent components interact.}
\label{fig:Triads}
\end{figure}

Common choices for the DSS
averaging operation, $\overline{\cdots}$, are temporal, spatial and ensemble means.
For models with zonal symmetry such as idealized geophysical or astrophysical flows
the zonal spatial average, defined as
$\overline{q}(\theta) \equiv \frac1{2\pi} \int_{0}^{2\pi} q(\theta,\phi) d\phi.$ is often the simplest and most natural.
Henceforth $\overline{\cdots}$ will be used to only denote such
a zonal average. Since the fluctuations $q^{\prime}$ with respect
to a zonal average now represent eddies that may be classified by zonal wavenumber, zonal CE2 allows for the
interaction between a mean flow and an eddy to produce an eddy, and also
for the interaction between two eddies to produce a mean flow
as shown in Figure \ref{fig:Triads}(a). 
An important virtue of the zonal average is that it reduces the dimension of each cumulant by one in the zonal direction,
reducing computational complexity.
The first cumulant depends only on the co-latitude,
$c(\vec{\Omega}_1) = c(\theta_1)$,
and the second cumulant depends only on the two co-latitudes of either
points and the difference in their longitudes,
$c(\vec{\Omega}_1,\vec{\Omega}_2)=c(\theta_1,\theta_2,\phi_2-\phi_1).$
Since $c(\vec{\Omega_1})$, $p(\vec{\Omega_1})$ and
$f(\theta_1)$ do not vary with longitude, the second term in Equation~(\ref{eq:c1}) vanishes.  

\subsection{Ensemble Average}
\label{sub:EnsembleAverage}
In systems where the eddy-eddy scattering processes dominate, or where non-zonal coherent structures
dominate, zonal CE2 can give an inaccurate reproduction
of the statistics \citep{tobias2013direct}.  An alternative version of the cumulant expansion
that does not entirely neglect these processes can be formulated by replacing the zonal average with
an ensemble average, to be denoted in the following with $\left\langle \cdots\right\rangle$.
Formally the triadic interactions that are retained in
ensemble CE2 are identical to those in zonal CE2 as shown in Figure \ref{fig:Triads}(b) except for the addition of a diagram with three mean fields. This addition reflects the fact that, for ensemble averaging,
the mean field may contain non-zonal coherent structures \citep{bakas2011structural,bakas2013emergence,bakas2014theory,Allawala:2016hx}
instead of only the zonal mean; the fluctuations about this average
represent incoherent perturbations, instead of eddies, that may interact.  
The coherent part of the flow, $c(\theta, \phi) = \langle \zeta(\theta, \phi) \rangle$,
may for instance consist of long-lived vortices rather than zonal jets \citep{kolmogorovForcing2017}
in which case departures from the zonal mean do not necessarily
constitute a fluctuation, and some of the eddy + eddy $\rightarrow$ eddy 
scattering processes (where eddies are defined as deviations from the {\it zonal} mean) are retained.  
On the other hand, because only the first two cumulants are retained,
ensemble-averaged CE2 is equivalent to the requirement that the probability distribution function 
be purely Gaussian -- an approximation
known to be poor for some flows such as isotropic and homogeneous turbulence.  

It is well known that for ergodic flows, the average over an infinite ensemble of realizations should equal a long-time average (in the limit of the averaging time going to infinity). For such flows one therefore expects that the statistics in ensemble CE2 typically
flow to a single fixed point --- this would be guaranteed if there were no closure approximation ---  and those for zonal CE2 to match this behaviour. However in flows where long-lived coherent structures such as vortices and jets play an important role \citep{frishman2017}, reducing the chaoticity of the flow, it is not true that the two versions of CE2 described here
should yield the same results. If ensemble CE2 is initialised with non-zonally symmetric initial conditions then we find that it oftn flows to a fixed point. By contrast zonal CE2 often does not flow to a fixed point; instead the statistics may
oscillate in time \citep{marston2014direct}.

\section{Proper Orthogonal Decomposition}
\label{sub:POD}

The zonal average second cumulant is a three-dimensional object, one dimension higher 
than that of the underlying dynamical fields;  for 
ensemble averages the second cumulant is of dimension four.  
This ``curse of dimensionality''\citep{bellman1957,bellman1961} can be tamed by application of Proper Orthogonal Decomposition
(POD) \citep{holmes1998turbulence,muld2012flow} directly to the low order statistics.  
Traditionally, POD  has been applied directly to the instantaneous master PDEs; instead, here we apply these methods directly to the statistical formulation and find improved performance.  Because the two operations of formulating DSS, and reducing dimensionality, do not commute, a given number of POD modes may better represent the DSS statistics than they would the full instantaneous dynamics. In the reduced basis the two-point function may be efficiently evolved forward in time without encountering the instabilities that plague DNS in POD bases \citep{Resseguier:2015tr}.  Here we illustrate how the procedure keeps dimensionality in check without significant loss of accuracy.  

A new basis of lower dimensionality that represents the first and second 
cumulants may be found by Schmidt decomposition of the zonally-averaged second moment:  
\begin{eqnarray}
\overline{\zeta(\vec{\Omega}_1)~ \zeta(\vec{\Omega}_2)} &=&  c(\vec{\Omega}_1, \vec{\Omega}_2) + c(\theta_1) c(\theta_2)
\nonumber \\
&=& \sum_i \lambda_i~ \phi_i(\vec{\Omega}_1)~ \phi_i(\vec{\Omega}_2)\ .
\label{Schmidt}
\end{eqnarray}
Here $\phi_i(\vec{\Omega}_1)$ is an eigenvector of the second moment with eigenvalue $\lambda_i$ 
that is both real and non-negative \citep{kraichnan1980realizability}.  Eq. \ref{Schmidt} can equivalently be expressed in the space of spherical harmonics where the second moment $m_{\ell \ell^{\prime} m} \equiv \zeta_{\ell m} \zeta_{\ell^\prime m}^*$ is block-diagonal in the zonal wavenumber $m$ and
\begin{equation}
m_{\ell \ell^{\prime} m} = c_{\ell \ell^{\prime} m} + c_{\ell} c_{\ell^{\prime}} \delta_{m,0} = \sum_{i} U_{\ell i}^{(m)} \lambda_{i}^{(m)} U^{\dagger (m)}_{i \ell^{\prime}}\ .
\label{unitary}
\end{equation}
Here $U_{\ell i}^{(m)}$ are unitary matrices ($\sum_\ell U^{\dagger (m)}_{i \ell} U_{\ell j}^{(m)} = \delta_{i j}$) 
that are composed of the eigenvectors of the second moments: 
$\sum_{\ell^\prime} m_{\ell \ell^\prime m} U^{(m)}_{\ell^\prime j} = \lambda_j U^{(m)}_{\ell j}$.  
The dimension of the EOMs for the cumulants may now be reduced by setting a 
cutoff for the eigenvalues, $\lambda_{c}$, and discarding all eigenvectors
with eigenvalues below this value.  The truncation generally
breaks conservation of angular momentum, energy, and enstrophy, but for driven and damped systems
this will generally not cause divergences as long as the truncation is not too severe.  Positivity of the second cumulant is maintained by periodically projecting out eigenvectors with negative eigenvalues.

\begin{figure}
\begin{centering}
\includegraphics[clip,width=1\columnwidth]{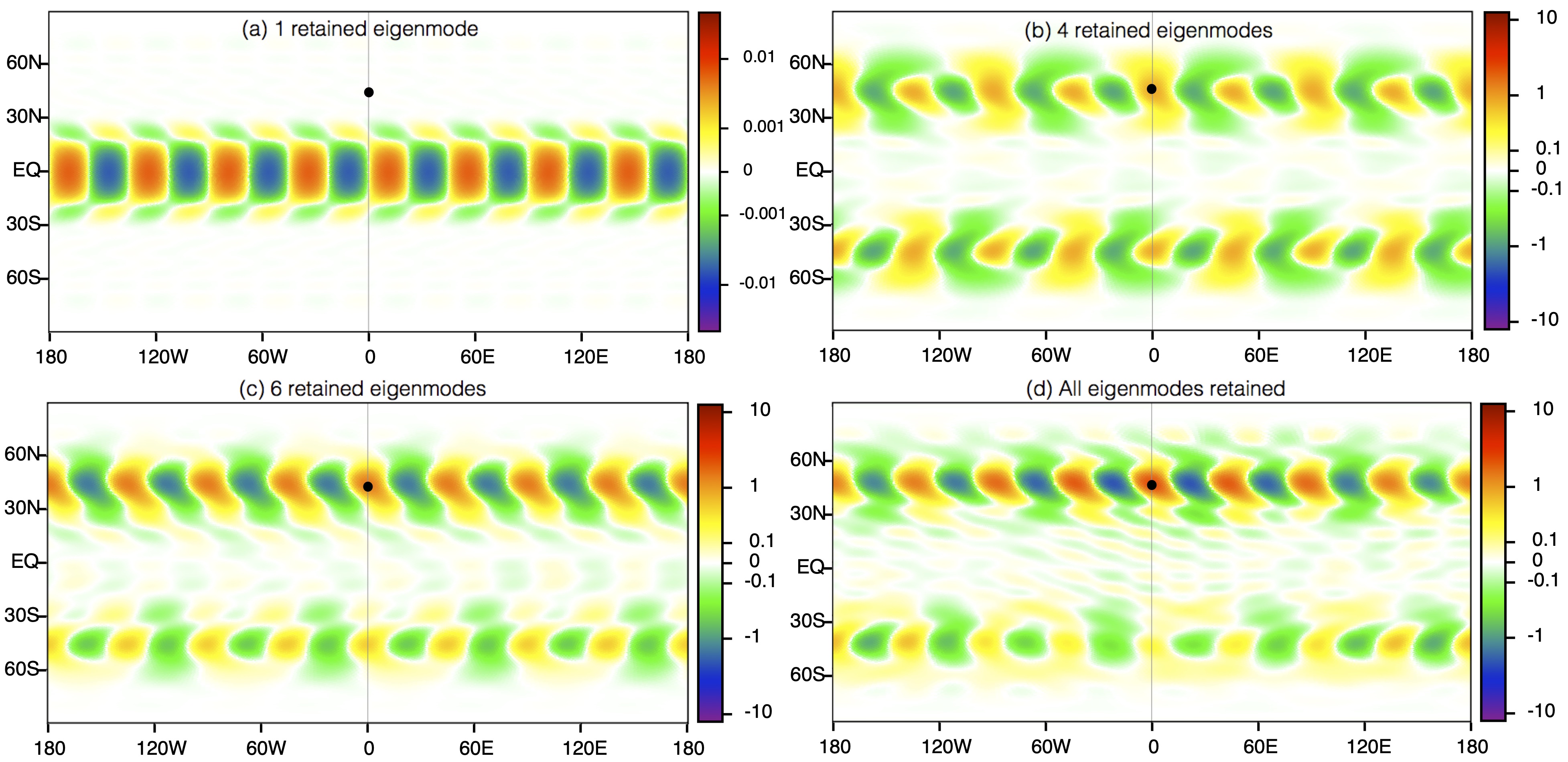}
\par\end{centering}
\caption{The reconstituted second cumulant of the stochastic jet with varying number of retained eigenmodes (out of a total of 441) as found by zonal CE2. 
The jet is spun up for $300$ days and then the second cumulant is time-averaged for a subsequent $700$ days.  The reference point (black dot) for one of the two coordinates appearing in the second cumulant, Equation~(\ref{cumulants}), is on
the central meridian ($\phi_2 = 0$) at co-latitude of $\theta_2 = 45^{\circ}$.  The 6 largest eigenvalues paired with their corresponding zonal wavenumber are: $(m,\ \lambda^{(m)}_i)$ = $(0,\ 19.43)$, $(8,\ 2.197)$, $(7,\ 2.062)$, $(4,\ 1.110)$, $(8,\ 0.838)$, and $(9,\ 0.676)$.}
\label{fig:SecondCumulantRestrictedEigenmodes}
\end{figure}

The zonally-averaged second moment is accumulated after spin-up to a statistical steady
state.  For each type of simulation (DNS, zonal CE2, or ensemble CE2) we first compute the second moment in the full basis, perform the POD decomposition, and implement the corresponding reduced-dimensionality version of CE2.  Figure \ref{fig:SecondCumulantRestrictedEigenmodes}, for instance, shows the second cumulant for the stochastically-forced jet (defined below) as calculated by zonal CE2 at  
different levels of truncation.  A severe truncation that retains only the six eigenvectors with largest eigenvalues is still able to reproduce
most features of the cumulant.   

\section{Tests}
\label{sec:Tests}

We first implement unreduced ensemble CE2 and zonal
CE2 on two idealized barotropic models
on a rotating unit sphere. After
comparing these two variants of CE2 against
the statistics gathered by DNS that are used as the reference truth, we  study the efficacy of dimensional
reduction of the two forms of DSS using the POD procedure 
outlined in Section \ref{sub:POD}.

\subsection{Point Jet}
\label{sub:DeterministicJet}

\begin{figure}
\includegraphics[clip,width=1\columnwidth]{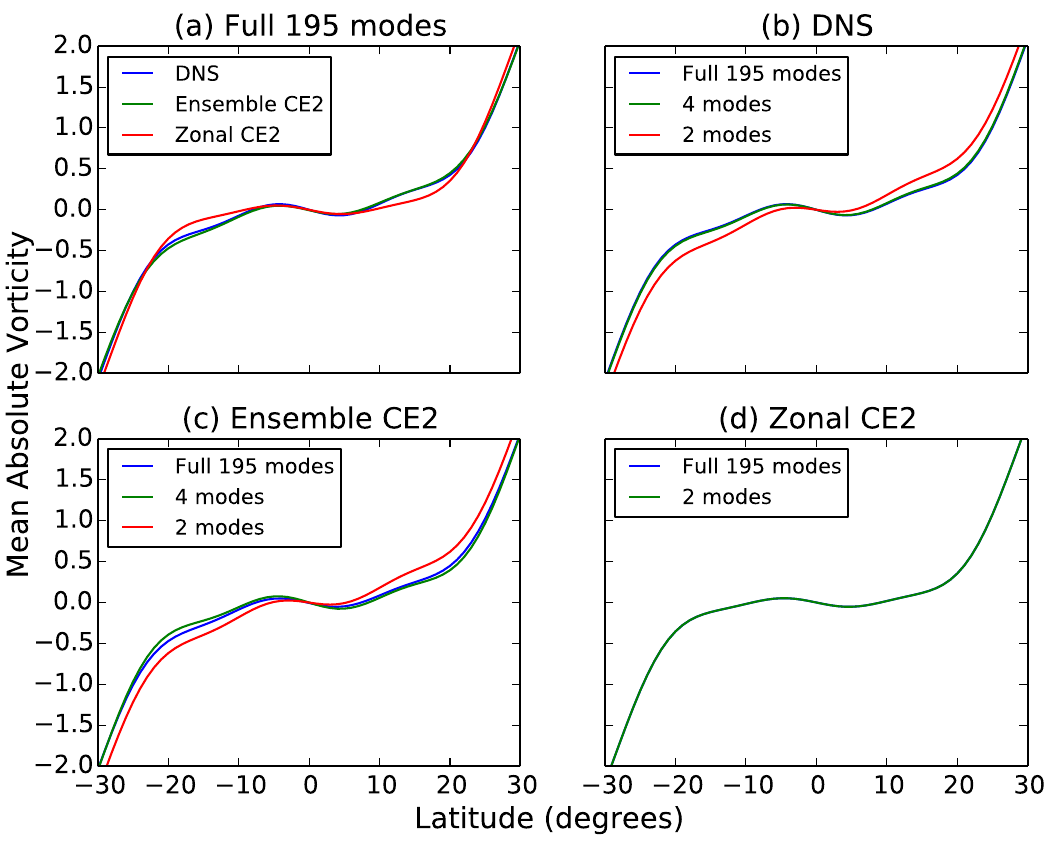}
\caption{(a) Zonal mean absolute vorticity $\overline{\zeta + f}$ 
of the point jet as a function
of latitude for DNS, ensemble CE2 and zonal
CE2. (b) -- (d) Comparison of all modes retained
against different levels of truncation for DNS, ensemble CE2
and zonal CE2 respectively. Each run is evolved until a statistically-steady
state is reached.
\label{fig:MeanAbsoluteVorticity}}
\end{figure}

\begin{figure}
\includegraphics[clip,width=1\columnwidth]{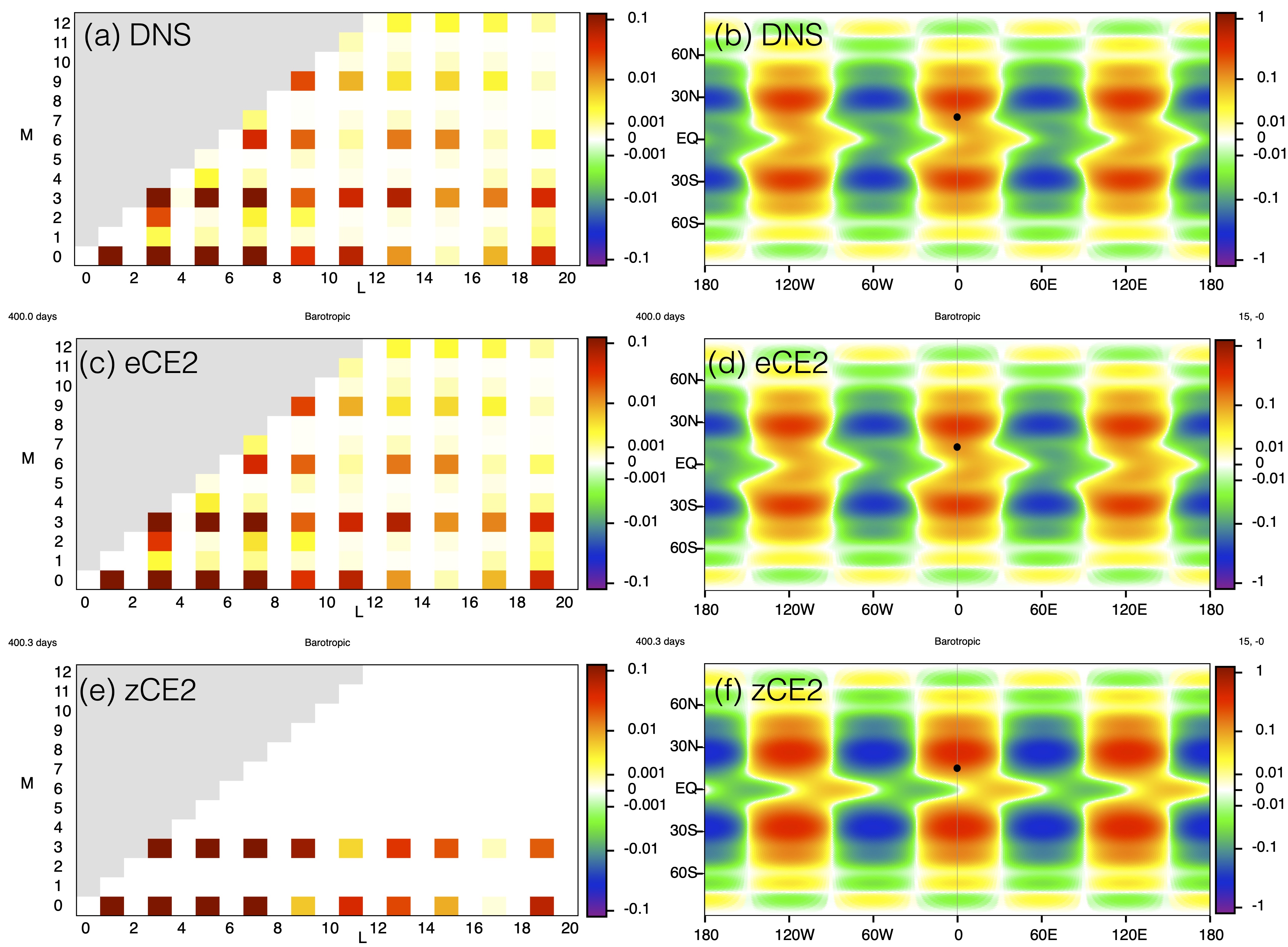}
\caption{Power spectra (left) and second cumulant of relative vorticity field (right)
of the point jet for (a-b) DNS, (c-d) ensemble CE2 
and (e-f) zonal CE2.  DNS is
evolved until $1000$ days with time-averaging over the last $500$ days. Ensemble
and zonal CE2 are evolved until $400$ days without time-averaging.
The reference point (black dot) for one of the two coordinates appearing in the second cumulant, Equation~(\ref{cumulants}), is on
the central meridian ($\phi_2 = 0$) at co-latitude of $\theta_2 = 75^{\circ}$. Ensemble
CE2 matches the statistics obtained by accumulation from DNS whereas zonal CE2 exhibits an artificial
symmetry of reflection about the equator not present in the statistics obtained from DNS.
\label{fig:PointJetSpectraCumulants}}
\end{figure}

We study a deterministic point jet that relaxes on a timescale $\tau$ toward an unstable
retrograde jet with meridional profile $\zeta_{jet}(\theta) = - \Xi \tanh((\pi / 2 - \theta)/{\Delta \theta})$. 
Parameters $\Xi = 0.6 \Omega$ and $\Delta \theta = 0.05$ are chosen to correspond with the jet 
previously examined in \citet{marston65conover}. 
The flow is damped and driven by the terms that appear in Equation~(\ref{eq:BarotropicEOM}), i.e.
\begin{equation}
L\left[\zeta\right] + F = 
\frac{\zeta_{jet} - \zeta}{\tau}\ 
\label{relaxation}
\end{equation}
with $\eta = 0$.
Spectral simulations are performed at a modest resolution of $0 \leq \ell \leq L$ and $|m| \leq min\{\ell, M\}$ with $L = 20$
and $M = 12$.  For a short relaxation time of $\tau=2$ days the
flow is dominated by critical-layer waves and eddies are suppressed by
the strong coupling to the fixed jet.  For a longer relaxation
time of $\tau=20$ days the flow is turbulent and well-mixed near the equator.  Since zonal
CE2 neglects eddy-eddy interactions it increasingly differs from DNS as $\tau$ grows. Here we
set $\tau=20$ days to demonstrate that ensemble-averaged CE2 captures
features of the non-zonal coherent structures that zonal average CE2
misses.  Figure \ref{fig:MeanAbsoluteVorticity}(a) shows that ensemble 
CE2 accurately reproduces the zonal mean absolute vorticity, whereas zonal CE2 overmixes
the absolute vorticity at low latitudes \citep{marston65conover}.  
Figure \ref{fig:PointJetSpectraCumulants} shows that whereas zonal
CE2 has power only in the zonal mean $m=0$ mode, and a single eddy of wavenumber $m=3$, the power spectrum
of ensemble CE2 matches that found by DNS.  DNS statistics are accumulated from time $500$
to $1000$ days after spin-up of  $500$ days.  Ensemble
CE2 reaches a statistical steady-state by $400$ days and no time-averaging
is required. Whereas the second cumulant of the vorticity field as determined by zonal CE2 
exhibits artificial reflection symmetry 
about the equator \citep{marston65conover}, ensemble CE2 does not have this 
defect; see Figure \ref{fig:PointJetSpectraCumulants}(b, d, f).  

\begin{figure}
\centering \includegraphics[clip,width=0.5\columnwidth]{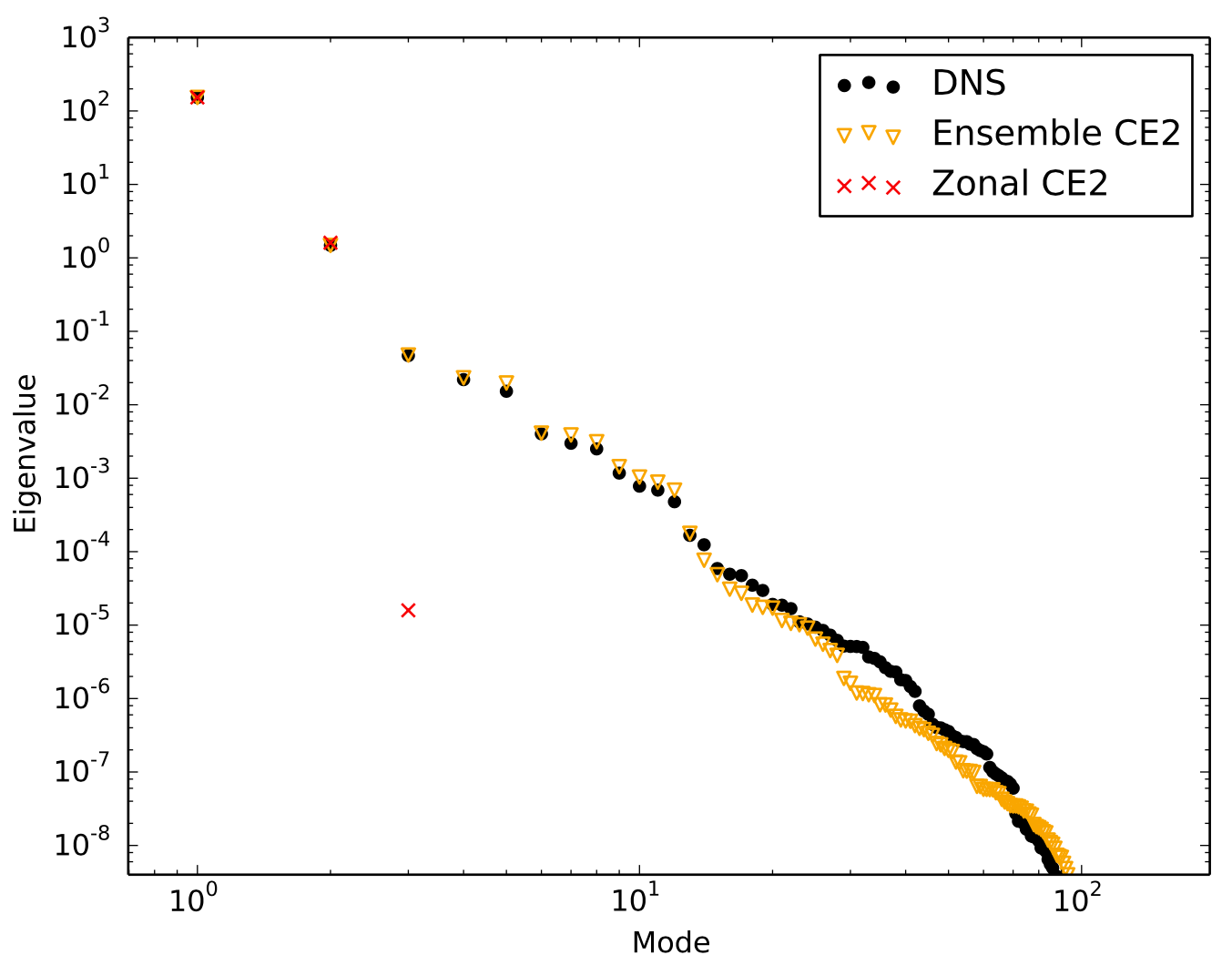} \vfill \centering \includegraphics[clip,width=0.5\columnwidth]{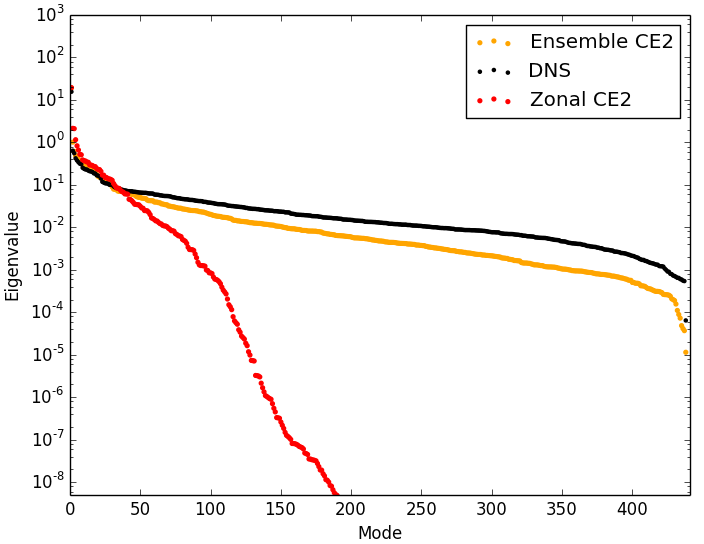}
\caption{(Top) Spectrum of eigenvalues of the second moment of the point jet
in descending order for DNS, ensemble CE2, and zonal CE2. The eigenvalues decay most quickly for zonal CE2; 
only zonal wavenumbers $m = 0$ and $m = 3$ have non-zero eigenvalues $\lambda_i^{(m)}$, consistent with the power spectra shown in Figure \ref{fig:PointJetSpectraCumulants}. 
(Bottom) Spectrum of eigenvalues for the stochastic jet.  The eigenvalues decay exponentially,
though less rapidly than for the point jet.
\label{fig:Eigenvalues}}
\end{figure}

We turn now to model reduction via POD truncation.
Figure \ref{fig:MeanAbsoluteVorticity}(b)-(d) show the first cumulants,
or zonal mean absolute vorticity, at different levels of truncation.
While POD offers a substantial dimensional reduction for DNS and ensemble
CE2 with slight loss of accuracy, extreme truncation of zonal CE2 down to just a few
modes is possible. This is consistent with the spectrum
of the eigenvalues of the second moment shown in Figure \ref{fig:Eigenvalues}(a).
The steep decay of the eigenvalues of zonal CE2 reflects the fact that 
only the $m = 3$ eddy has power.  

\subsection{Stochastic Jet}
\label{sub:StochasticJet}

Jets that form spontaneously in the presence of rotation and small scale random forcing 
provide another, perhaps more stringent, test of DSS.  An idealized barotropic model that has been much studied \citep{Farrell:2007fq,tobias2011astrophysical,tobias2013direct,Constantinou:2013fh,marston2014direct} is governed by:
\begin{equation}
L\left[\zeta\right] = - \left[\kappa-\nu_3(\nabla^2 + 2)\nabla^4\right]\zeta,
\end{equation}
with $F = 0$.  We examine this model for the set of parameters used in \citep{marston2014direct};
friction $\kappa=0.02$ with hyperviscosity $\nu_3$ set such that the mode at the smallest
length scale decays at a rate of 1. The covariance of the Gaussian white noise of the modes that are forced stochastically is $\Gamma_{\ell \ell^\prime m} = 0.1 \delta_{\ell \ell^\prime}$ for $8 \leq \ell \leq 12$ and $8 \leq |m| \leq \ell$,
and set the stochastic renewal time to be $\tau_r = 0.1$.
For this experiment, the forcing is concentrated at low latitudes, enabling an evaluation of the ability of different 
DSS approximations to convey angular momentum towards the poles \citep{marston2014direct,Marston:2016ff}.
Spectral simulations are performed at a resolution of $0\leq l\leq L$ and $|m|\leq min\{l,M\}$ with $L = 30$
for $M = 20$.  After a spin-up time of $300$ days, statistics are accumulated for a further $700$ days.  

\begin{figure}
\includegraphics[clip,width=1.0\columnwidth]{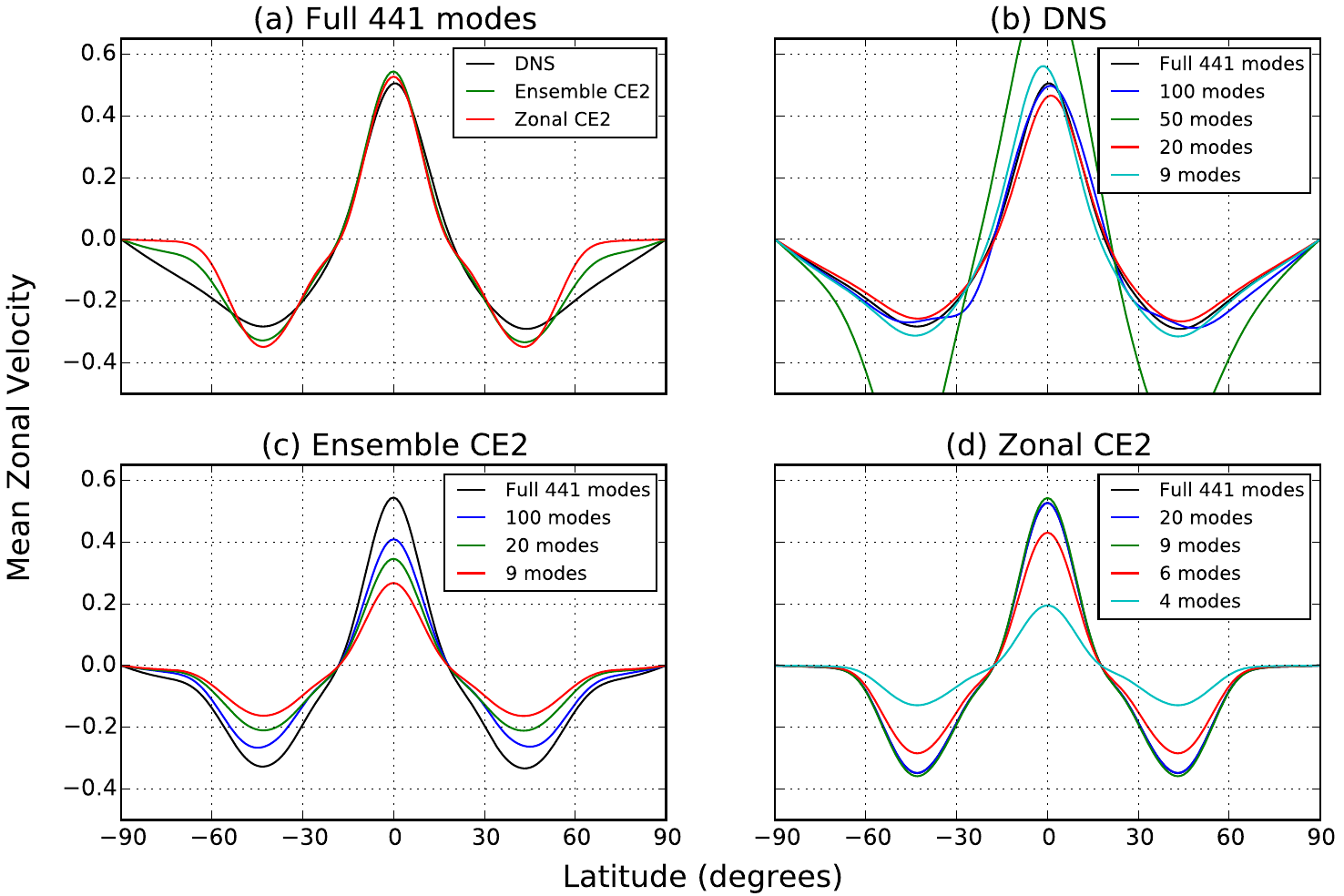}
\caption{(a) Comparison of the zonal mean zonal velocity of the stochastically-driven jet as a function of latitude
for DNS, ensemble CE2 and zonal CE2. Ensemble CE2 matches DNS better
than zonal CE2 because it retains some of the eddy-eddy scattering process
which transports eddy angular momentum poleward. (b)-(d) Comparison of the
zonal mean zonal velocities of the non-truncated system against different
levels of truncation for DNS, ensemble CE2 and zonal CE2 respectively. 
\label{fig:ZonalMeanVelocity}}
\end{figure}

\begin{figure}
\centering \includegraphics[clip,width=0.7\columnwidth]{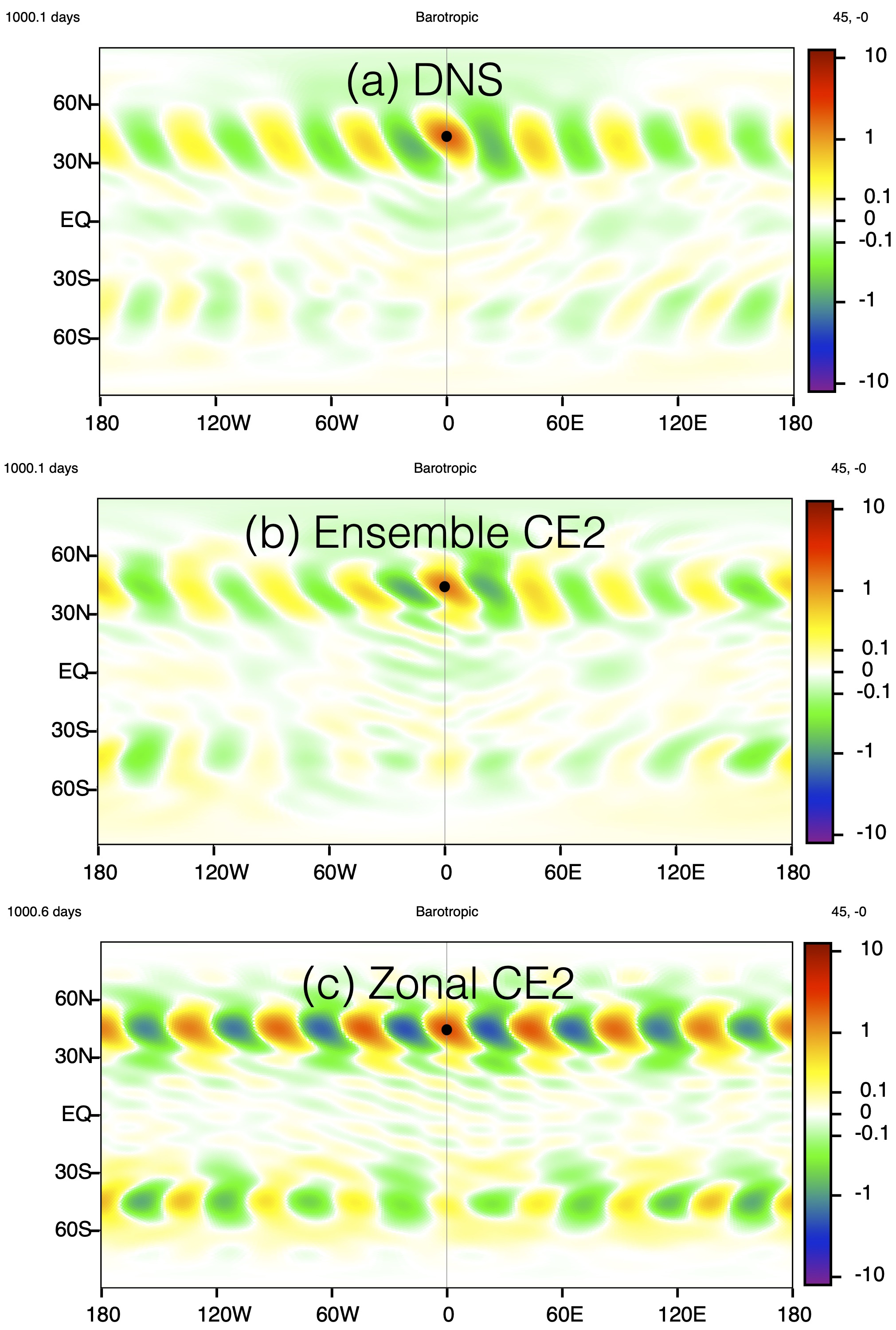}
\caption{Second cumulant of relative vorticity field of the stochastic jet for (a) DNS, 
(b) ensemble CE2, and (c) zonal CE2.  The reference point for the second cumulant (black dot) is positioned
along the central meridian at a co-latitude of $45^{\circ}$.  Zonal CE2 shows
exaggerated coherent waves in comparison to DNS and ensemble CE2.
\label{fig:StochasticJet}}
\end{figure}

Figure \ref{fig:ZonalMeanVelocity}(a) compares the zonal mean zonal velocity as a function
of latitude for ensemble CE2, zonal CE2, and DNS.  Owing to the neglect of eddy-eddy scattering in zonal CE2, 
there is no mechanism to transport eddy angular momentum
from the equator, where the forcing is concentrated, towards the poles, and the method underestimates the
mean zonal velocity at high latitudes.  Ensemble CE2 does somewhat better, but the 
match at high latitudes remains poor.  (Higher-order closures do much better -- see \citet{marston2014direct} -- but for simplicity we do not study them here.)  The second vorticity cumulant
is shown in Figure \ref{fig:StochasticJet}(b,d,f).  Vorticity correlations are non-local
in space, and zonal CE2 exaggerates the range of the correlations because
the waves are coherent, again owing to the lack of eddy-eddy scattering.   
Ensemble CE2 more closely matches DNS.

\begin{figure}
\begin{centering}
\includegraphics[clip,width=1\columnwidth]{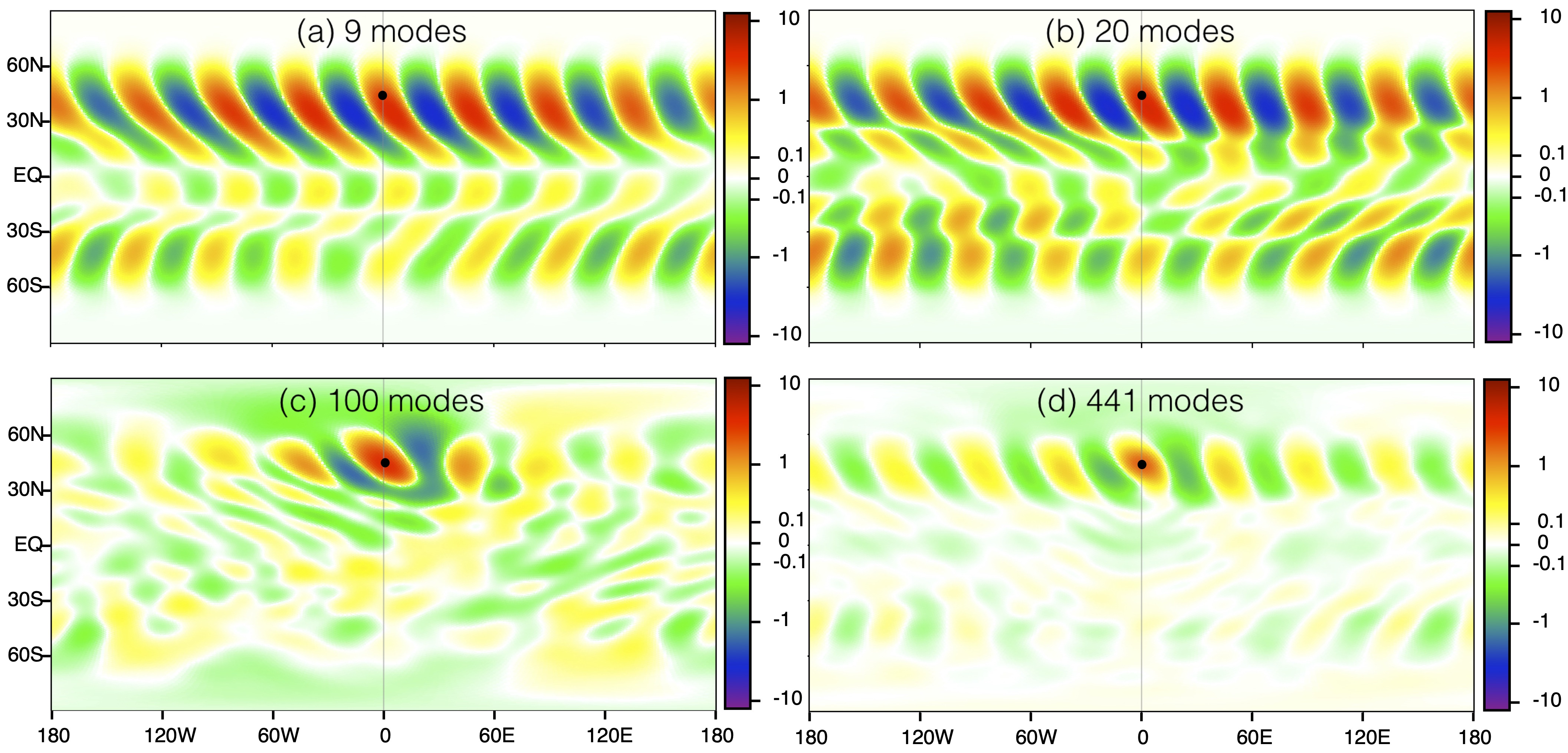}
\par\end{centering}
\caption{Comparison of the second vorticity cumulants of the stochastic jet
in the statistical steady state of DNS for different truncation levels.  The 
non-truncated statistics are shown in panel (d). 
\label{fig:2ndCumulantsDNS}}
\end{figure}

\begin{figure}
\begin{centering}
\includegraphics[clip,width=1\columnwidth]{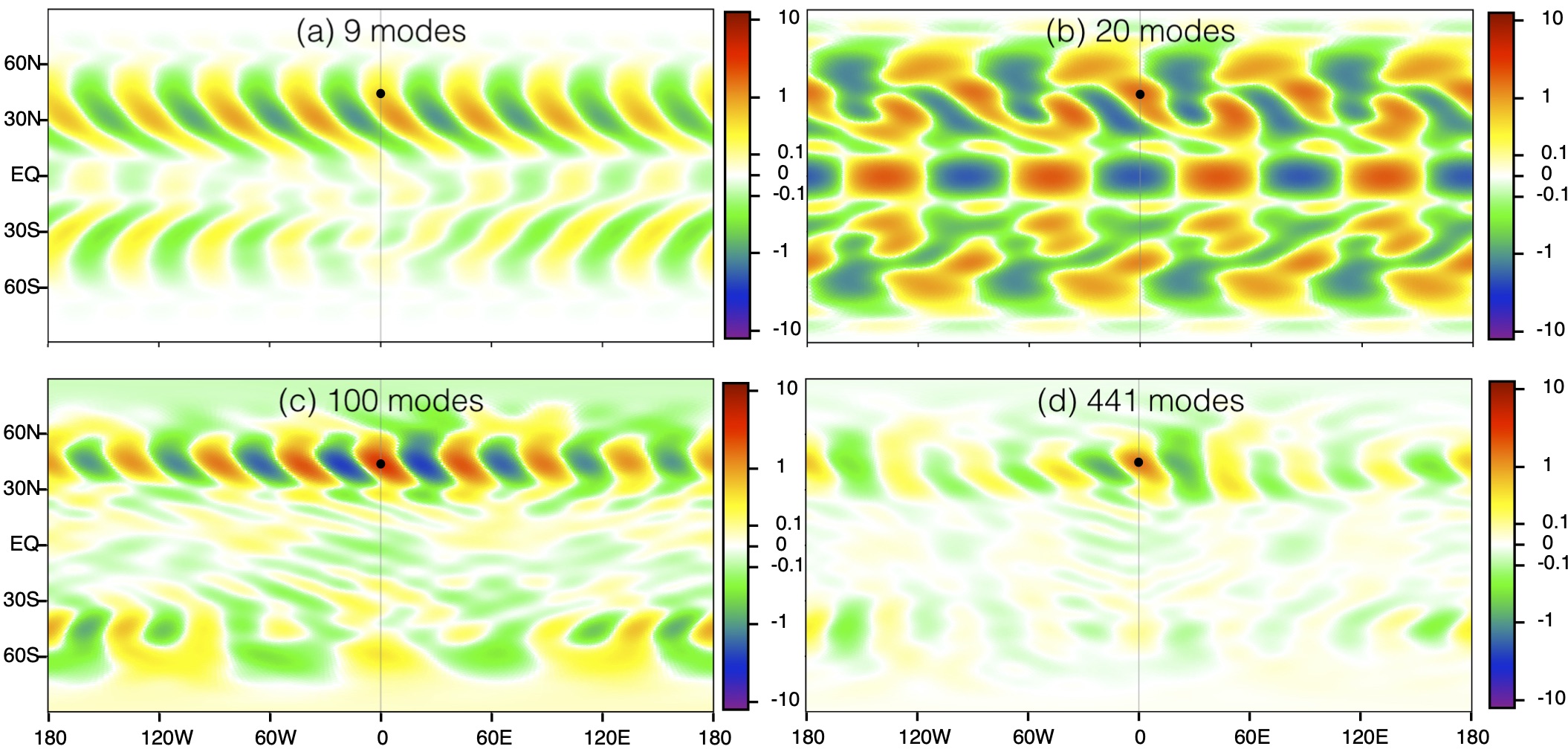}
\par\end{centering}
\caption{Comparison of the second vorticity cumulants of the stochastic jet
in the statistical steady state of ensemble CE2 for different truncation levels.  The 
non-truncated statistics are shown in panel (d).}.
\label{fig:2ndCumulantseCE2}
\end{figure}

We now examine the reduction in dimensionality of DSS by POD. Figure \ref{fig:ZonalMeanVelocity}(b)-(d)
shows the zonal mean zonal velocity as a function of latitude at different
levels of truncation for DNS, ensemble CE2 and zonal CE2.  It is apparent that both types of CE2 are better suited to the POD method than DNS.
Zonal CE2 in particular allows for a more severe truncation compared with ensemble CE2
and DNS, a fact that can again be explained by the spectrum of eigenvalues shown in Figure \ref{fig:Eigenvalues}(b).  
The eigenvalues decay more slowly for the stochastically-forced jet than for the point jet.  
It can be seen that statistics accumulated from DNS do not  converge monotonically in truncation level toward those of the full simulation; 
those obtained from the cumulant expansions are better behaved.  
Convergence of the second cumulant with increasing
number of retained modes is also evident; see Figures \ref{fig:2ndCumulantsDNS} and \ref{fig:2ndCumulantseCE2}.

The number of operations required for a POD reduced CE2 time step scales with the size of the
retained basis $N_{ret}$ as $\mathcal{O}(N_{ret}^3)$.
Therefore a dimensional reduction from $N_{ret}=441$ to $N_{ret}=13$
would naively equate to a speed up of order $10^4$.  In the case of unreduced zonal CE2,
however, zonal symmetry may be exploited to decrease the number of operations per time step \citep{marston2014direct} to 
$\mathcal{O}(L^{3}M) << \mathcal{O}((LM)^3)$.  Nevertheless, the speed up is 
considerable.   A machine with a 2.6 GHz quad-core Intel Core i7 processor runs $28 \times$ faster for reduced zonal CE2 
with $N_{ret}=13$ in comparison to full unreduced zonal CE2, with little loss in accuracy. 

\section{Continuation In Parameter Space}
\label{sec:Continuation}

The test problems studied in the preceding section show that it is possible to implement DSS in a subspace of reduced dimensionality, supporting the idea that relatively few modes are required for a description of the low-order statistics.  The reduced statistical simulations still, however, require a full resolution training run for POD.  Thus, from a practical point of view, the reduced simulations do not offer a speed-up.  In order to address this,  we show here that it is possible to fix the reduced basis obtained from a single training run, alter model parameters, and still obtain good agreement with both DNS and CE2 in a reduced basis. Thus, the reduced basis obtained from one training run can be used to perform DSS for a range of parameters.

\begin{figure}
\centerline{\includegraphics[clip,width=0.8\columnwidth]{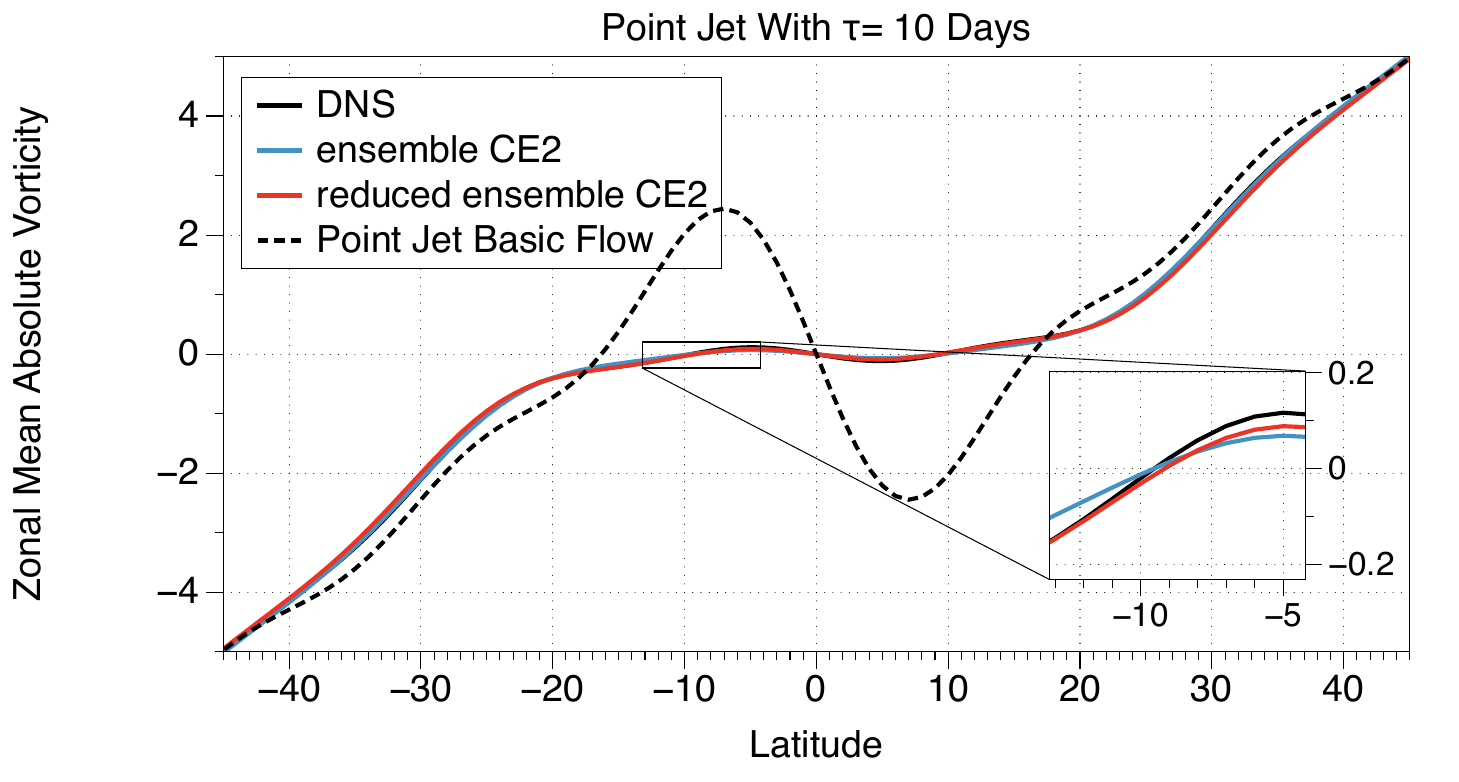}}
\centerline{\includegraphics[clip,width=0.8\columnwidth]{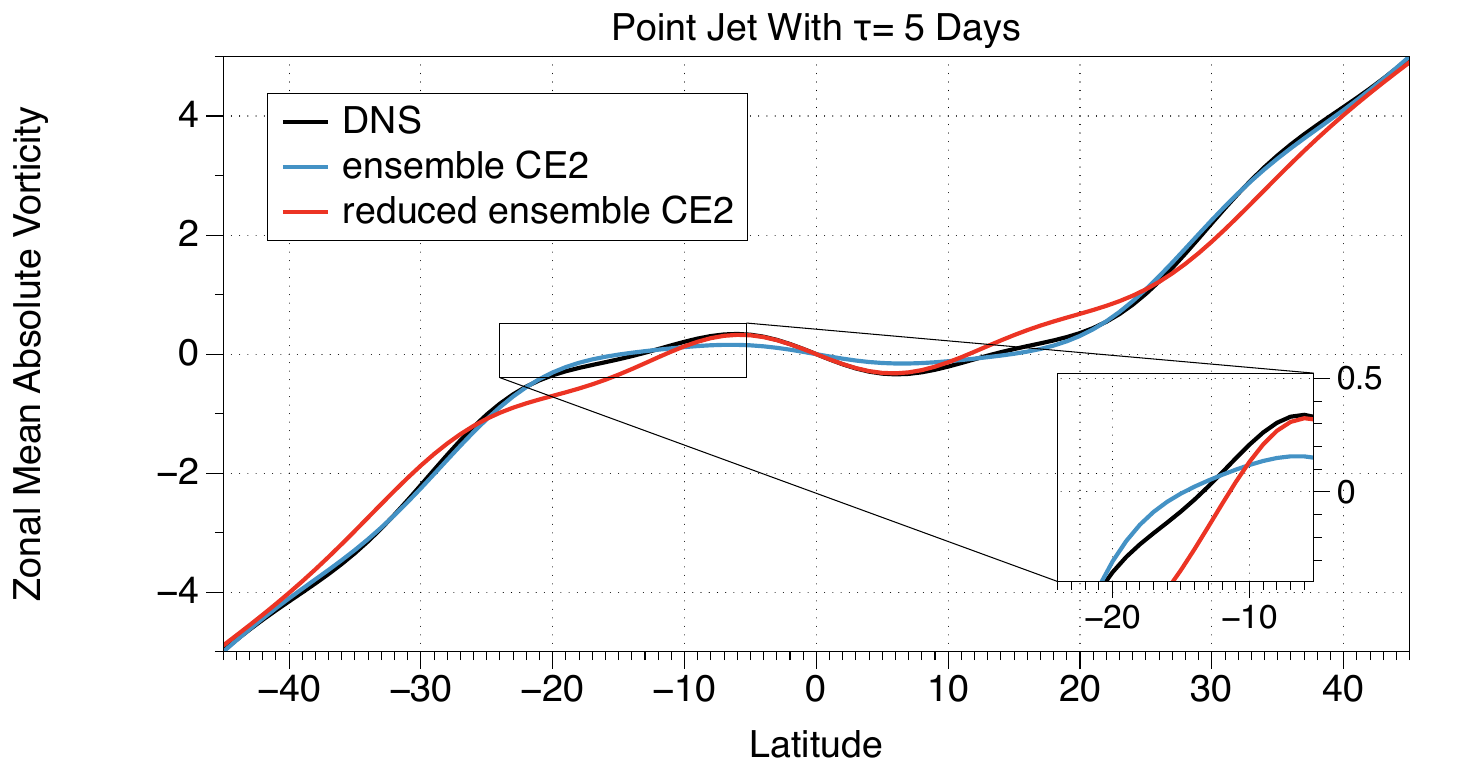}}
\caption{Zonal mean absolute vorticity $\overline{\zeta + f}$, 
of the point jet as a function of latitude for DNS, full ensemble CE2 and reduced ensemble CE2 for different jet relaxation times $\tau$ (see Eq. \ref{relaxation}).  The reduced basis is fixed and obtained by POD from ensemble CE2 for a single training run on the jet with $\tau = 20$ days.  40 modes are retained.  Top: The point jet with $\tau = 10$ days.  The basic absolute vorticity of the point jet, $\zeta_{jet} + f$, is shown for comparison; the region around the equator is well mixed by the eddies.  Bottom: The point jet $\tau = 5$ days.  As parameter $\tau$ is continued away from the training run with $\tau = 20$ the agreement between reduced ensemble CE2 with full ensemble CE2 and DNS deteriorates.  
\label{fig:Continuation}}
\end{figure}

We illustrate the continuation by changing the relaxation time of the point jet. This parameter changes the relative degree of turbulence to mean flow in the jet and is related to the Kubo number, which is a measure of the degree of applicability of the quasilinear approximation \citep{Marston:2016ff}. We utilize the $\tau = 20$ days jet as a training run for POD on the ensemble CE2 solution, and calculate the reduced basis for this parameter set.  Figure \ref{fig:Continuation} shows the (zonally averaged) first cumulant for ensemble CE2 in this reduced basis of $40$ retained modes for $\tau = 10$ and $5$ days.  Note that the dynamics (and indeed the statistics) does change significantly when this parameter is altered. For $\tau=10$ days the agreement is exceptionally good. However, as the parameter $\tau$ is moved further away from the training run to $\tau = 5$ days the agreement between reduced ensemble CE2 with full ensemble CE2 and DNS worsens as expected. Even here, qualitative agreement is retained.  

\begin{figure}
\centerline{\includegraphics[clip,width=0.8\columnwidth]{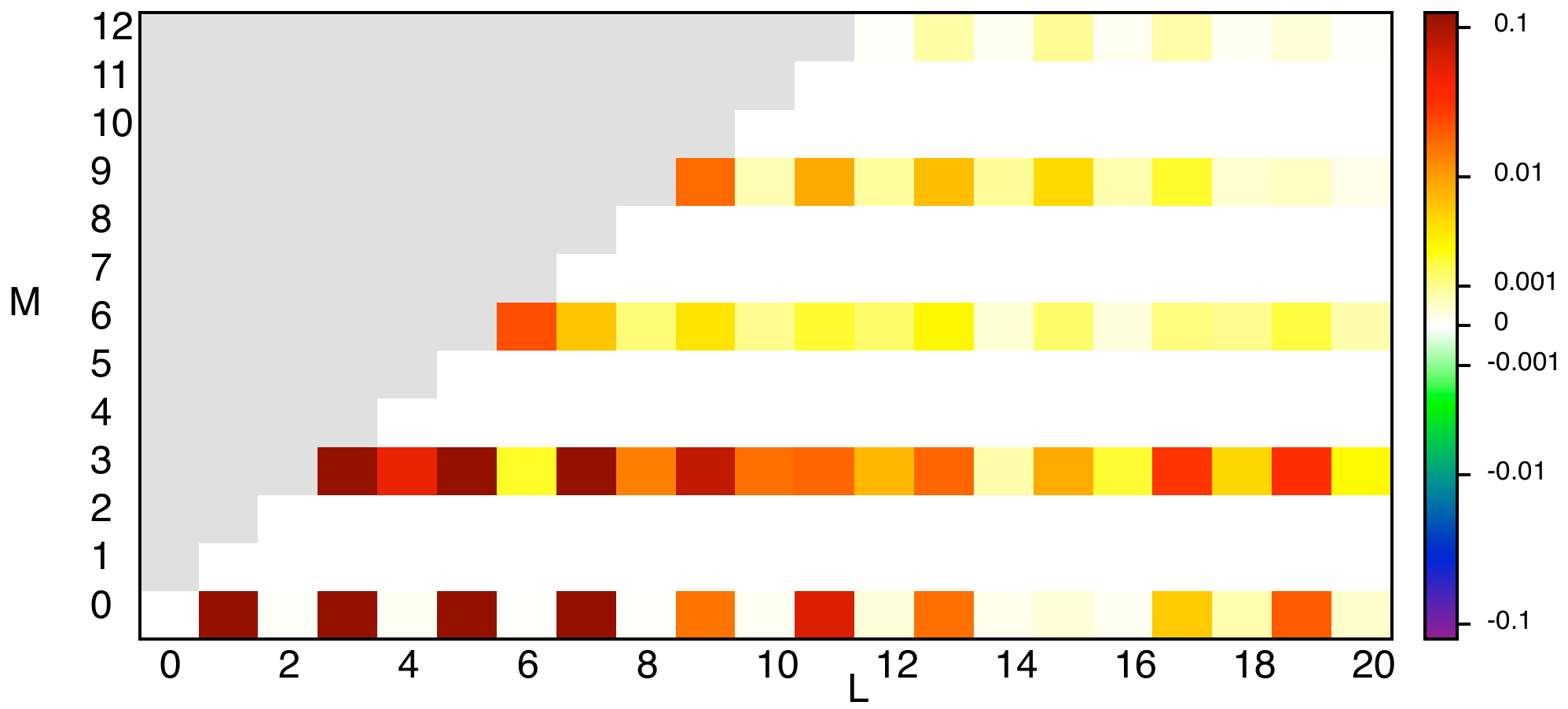}}
\centerline{\includegraphics[clip,width=0.8\columnwidth]{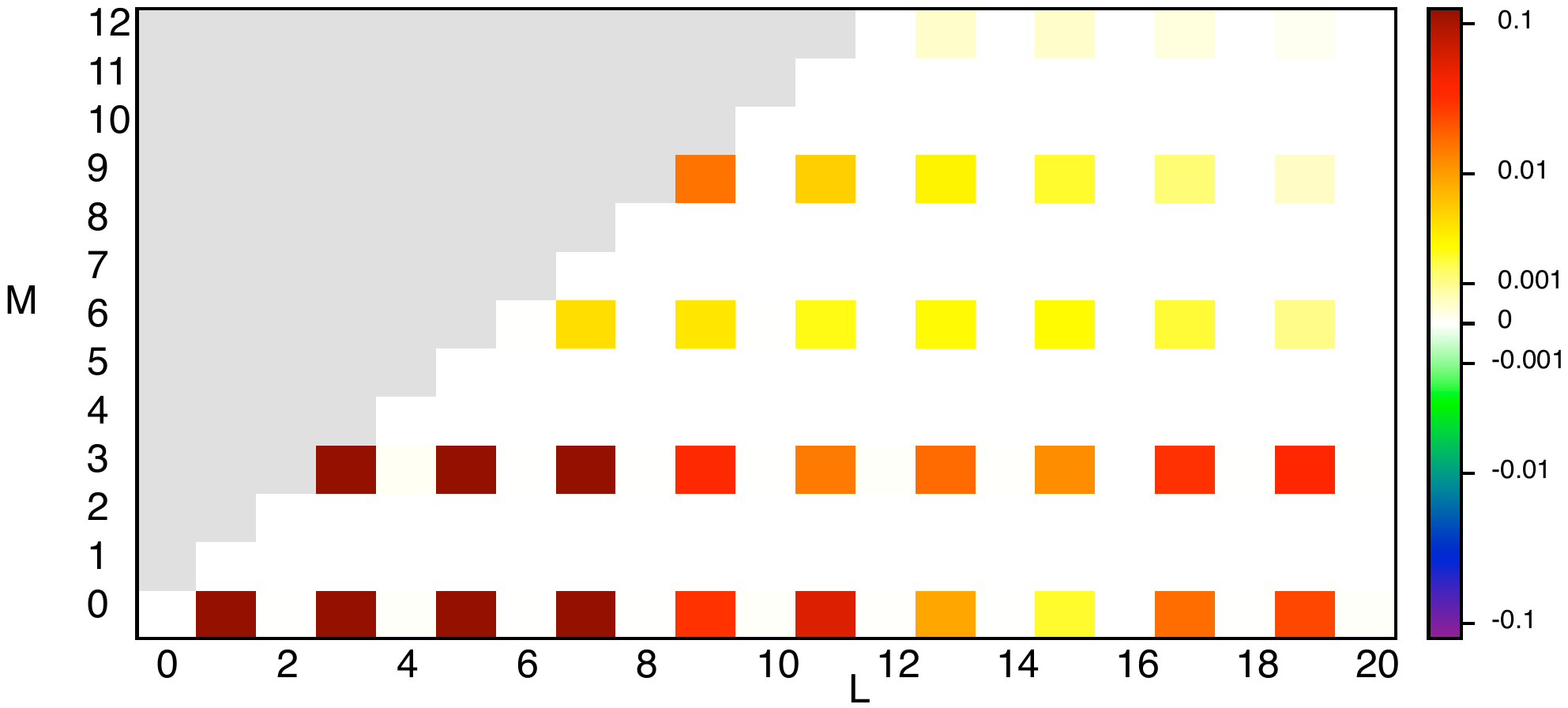}}
\caption{Power spectra for the $\tau = 10$ day jet.  Top:  ensemble CE2 with the full basis.  Bottom: reduced ensemble CE2 with 40 retained modes, with the fixed basis determined by training on the $\tau = 20$ day jet. 
\label{fig:ContinuedPower}}
\end{figure}

Comparison of power spectra in Figure \ref{fig:ContinuedPower} shows that second-order statistics also continue to show qualitative agreement at $\tau=10$ days. The reduction in dimensionality simplifies the spectrum of the reduced ensemble CE2 spectra in comparison with the full-resolution simulation as expected.   
This initial exploration of parameter continuation shows a promising direction for future research.  

\section{Discussion and conclusion}
\label{sec:Conclusion}

Proper Orthogonal Decomposition (POD) can be used to reduce the
dimensionality of second order cumulant expansions (CE2) by discarding modes that are unimportant
for the low-order statistics.  For the idealized models that we studied, the first and second cumulants can be accurately reproduced 
with relatively few modes, permitting a substantial reduction in dimensionality, and an increase in computational
speed. This is particularly true for zonal CE2.  The degree of truncation and hence computational saving that can be made while maintaining accuracy depends on the specifics of the system, as demonstrated by our results for the two different test problems.

It would be interesting to explore dimensional reduction
of Direct Statistical Simulation (DSS) for more realistic models such as those explored in \citet{ait2016cumulant}.  Dimensional reduction by POD has been tested in a quasilinear model of the ocean boundary layer \citep{Skitka:2020}.
It would also be interesting to explore whether or not a dimensional reduction algorithm could be 
constructed that acts dynamically on DSS, bypassing the step of first acquiring statistics
for the full non-truncated problem, as in the current work.  With such an advance it may be possible for
DSS to access regimes that cannot be reached by DNS.  We note that CE2 by itself has already been used for a dynamo problem to access lower magnetic Prandtl number than is possible by DNS \citep{Squire:2015ia}.   As reduced-dimensionality DSS is able to quickly describe some fluid dynamical systems via parameter continuation of the low-order statistics, the combination of POD with cumulant expansions offers a good prospect for other simulations to reach beyond DNS.

\section*{Declaration of Interests} The authors report no conflict of interest.

\section*{Acknowledgements}

Supported in part by NSF DMR-1306806 and by a grant from the Simons Foundation (Grant number 662962, GF).  SMT would also like to acknowledge support of  funding from the European Research Council (ERC) under the European Unions Horizon 2020 research and innovation program (grant agreement no. D5S-DLV-786780).  We thank Greg Chini and Joe Skitka for useful discussions.

\end{document}